\documentclass[aps,prb,twocolumn]{revtex4}
\usepackage{amssymb,lipsum}
\usepackage{graphicx,amsmath}
\setcitestyle{numbers,square}

\newcommand{\bea}{\begin{eqnarray}}
\newcommand{\eea}{\end{eqnarray}}
\newcommand{\beq}{\begin{equation}}
\newcommand{\eeq}{\end{equation}}
\newcommand{\benu}{\begin{enumerate}}
\newcommand{\enu}{\end{enumerate}}

\newcommand{\om}{\omega}
\newcommand{\Om}{\Omega}
\newcommand{\ep}{\epsilon}
\newcommand{\ta}{\tau}
\newcommand{\si}{\sigma}

\newcommand{\ham}{\mathcal{H}}

\newcommand{\bk}{{\bf k}}

\newcommand{\cda}{c^{\dagger}}
\newcommand{\bda}{b^{\dagger}}
\newcommand{\tp}{t^{\prime}}

\usepackage{braket}

\begin{document}

\title{
 Microscopic description of displacive coherent phonons}

\author{M. Lakehal and I. Paul}

\affiliation{
Laboratoire Mat\'{e}riaux et Ph\'{e}nom\`{e}nes Quantiques, Universit\'{e} Paris Diderot-Paris 7 \& CNRS,
UMR 7162, 75205 Paris, France
}

\begin{abstract}

We develop a Hamiltonian-based microscopic description of laser pump induced displacive coherent phonons.
The theory captures the feedback of the phonon excitation upon the electronic fluid, which is missing in
the state-of-the-art phenomenological formulation. We show that this feedback leads to chirping at short
time scales, even if the phonon motion is harmonic. At long times this feedback appears as a finite phase
in the oscillatory signal. We apply the theory to BaFe$_2$As$_2$, explain the origin of the phase in the
oscillatory signal reported in recent experiments, and we predict that the system will exhibit red-shifted
chirping at larger fluence. Our theory also opens the possibility to extract equilibrium information from
coherent phonon dynamics.

\end{abstract}

\date{\today}

\maketitle
%%%%%%%%%%%%%%%%%%%%%%%%%%%%%%%%%%%%%%%%%%%%%%%%%%%%%%%%%%%%%%%%%%%%%%%%%%

\section{Introduction}
\begin{figure}[!!b]
\begin{center}
 \includegraphics[width=1.0\linewidth,trim=0 0 0 0]{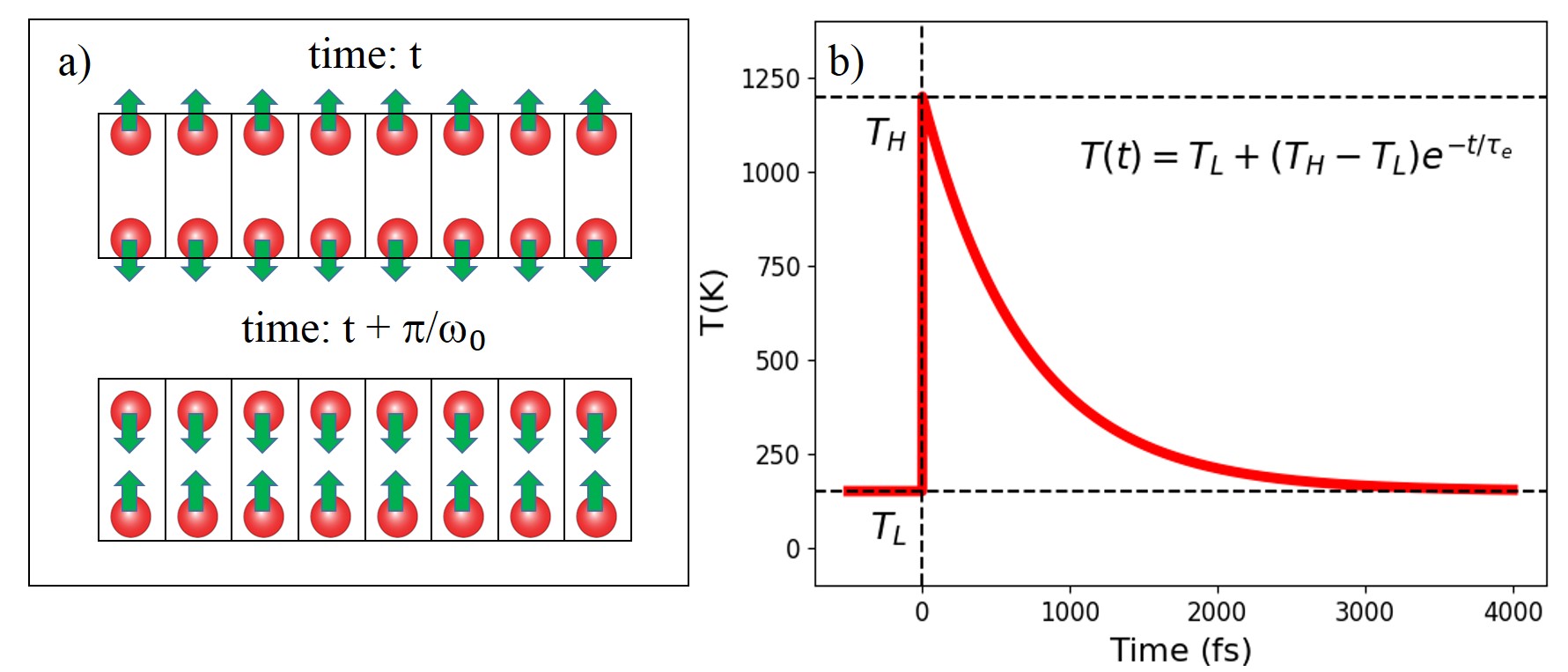}
\caption{
(color online) (a) Sketch of an $A_{1g}$ coherent phonon motion in a two-atom (red balls) unit cell. A macroscopic
number of atoms oscillate with \emph{identical} phase and frequency $\om_0$. Green arrows indicate the
instantaneous velocities at two instants. The motion preserves the point group symmetry.
(b)
The effect of the laser pump is idealized as a temperature quench from a measured base temperature $T_L$ to a high
temperature $T_H$ over a short time set to zero, and the subsequent relaxation of temperature over a time-scale $\tau_e$.
In the theory, $(T_H, \tau_e)$ are phenomenological parameters (see text). The temperature and time scales are
representative.
}
\label{fig1}
\end{center}
\end{figure}

The development of femtosecond laser pumps has led to new probes of complex metals whereby systems are
driven out of equilibrium with the aim to study their relaxation dynamics~\cite{giannetti2016,bovensiepen2007,smallwood2016}.
Simultaneously, pump-probe
setups allow the fascinating possibility to study phenomena that have no analog in equilibrium physics,
such as the transient excitation of coherent optical phonons~\cite{silvestri1985,chen1990,cho1990,ishioka2009}.
A ``coherent'' phonon is excited when the
relevant atoms of the crystalline solid, which are macroscopic in number,
vibrate with \emph{identical} frequency and phase [see Fig.~\ref{fig1}(a)]. This is to be contrasted with
incoherent motion triggered by quantum and thermal fluctuations in equilibrium where, from atom to atom,
the frequencies and phases are uncorrelated.
More recently it has been recognized that the
physics of Floquet dynamics can be made experimentally accessible via coherent phonon
excitations~\cite{hubener2018,oka2018}.

Experimentally, a typical signature of a
coherent phonon excitation is an oscillatory signal on a decaying background
in time-resolved spectroscopic probes such as x-ray spectroscopy, photoemission and reflectivity
measurements.
Coherent phonons have been studied in a variety of materials that include
semiconductors~\cite{hunsche1995,riffe2007,bothschafter2013},
semimetals~\cite{hase1996,decamp2001,hase2002,misochko2004,fritz2007,johnson2008,papalazarou2012},
transition metals~\cite{hase2005},
Cu-based~\cite{chwalek1990,albrecht1992,bozovic2004,novelli2017}
and the Fe-based~\cite{mansart2009,mansart2010,torchinsky2011,kim2012,avigo2013,yang2014,rettig2015,gerber2015,gerber2017}
high temperature superconductors,
charge density wave systems~\cite{kenji1998,demsar1999,toda2004,schaefer2014,schmitt2011},
as well as Mott~\cite{perfetti2008,mansart-mott-2010,mankowsky2015,lee2017} and
topological~\cite{a-q-wu2008,kamaraju2010,misochko2015}  insulators.

On the theory side, this phenomenon is usually described either as displacive excitaion of coherent phonons
(DECP)~\cite{zeiger1992,kuznetsov1994} or as impulsive stimulated Raman scattering (ISRS)~\cite{garrett1996,merlin1997}.
In the former mechanism
photoexcitation leads to a shift in the equilibrium position of the phonon~\cite{zeiger1992,kuznetsov1994},
while in the latter the electromagnetic
radiation provides a short impulsive force to the atoms~\cite{garrett1996,merlin1997}.
Note, if the photoexcitation does not involve crossing phase boundaries,
then typically only the fully symmetric Raman $A_{1g}$ phonon is excited in DECP.
It has been argued that in absorbing medium these two mechanisms
are not distinct~\cite{stevens2002}.
Using the above concepts, first-principles calculations have been successfully applied to understand coherent phonon
dynamics in a variety of systems~\cite{mazin1994,tangney1999,tangney2002,murray2005,shinohara2010}.

The purpose of this work is to develop, within the conceptual framework of DECP, a microscopic Hamiltonian-based
description of coherent phonons in an environment where the timescale for the photoexcited carriers to thermalize
is rather short, such as a metal with gapless charge excitations.
Here we focus on coherent phonon excitation driven by laser heating of carriers,
a phenomena which is relevant experimentally, but which has received less attention theoretically.
As we show below, the
microscopic formulation provides a better treatment of electron-phonon interaction compared to the phenomenological
model that is currently used to analyze experimental data~\cite{zeiger1992}.
In particular, our theory captures how the coherent phonon excitation
modifies the electronic fluid, and how this modification feeds back on the coherent phonon dynamics.
%====================
%====================

The main advances of our work compared to the phenomenological theory of Zeiger \emph{²}~\cite{zeiger1992}
are the following.
(i) Including the lattice feedback effect leads to a richer description of the dynamics. In particular, we show
that at short time scales this leads to
\emph{chirping} or temporal variations of the oscillation frequency, while staying within a harmonic description of
the coherent phonons. On the other hand, at long times the feedback leads to a finite phase in the oscillatory signal.
The origin of this phase is distinct from that in the phenomenological DECP theory~\cite{zeiger1992}, and it is likely
to be dominant quantitatively. Importantly,
the theory \emph{predicts} that the sign of the phase is determined by whether the chirping is red or blue shifted.
(ii) A Hamiltonian formulation opens the possibility of extracting microscopic equilibrium information from coherent phonon studies.
(iii) The microscopic formulation can be refined systematically using methods of many-body to deal with
various interaction effects.

The paper is organized as follows. In Sec.~\ref{Model}, we introduce the microscopic model,
we discuss the rationale for treating the effect of the pump as a quench of the electronic temperature,
and we derive the equation of motion of the coherent phonon using Heisenberg equation of motion.
In Sec.~\ref{Sec_results}, we solve the above equation, and we discuss our main results,
emphasizing the new physics introduced by taking into account the feedback of the lattice.
In Sec.~\ref{Ba_model}, we apply the theory to BaFe$_2$As$_2$ and we show that the data from a recent time resolved
x-ray study can be successfully described by our theory, using a more constrained fit.
We conclude in Sec. \ref{conclusion}.

\section{Model \& Formalism}\label{Model}
We consider a multiorbital electronic system interacting with a zero wavevector uniform $A_{1g}$ phonon mode.
It is described by the Hamiltonian
\begin{align}
\label{eq:ham}
\ham &= \sum_{\bk, a, b, \si} \left[ \ep(\bk)_{a b} - \mu \delta_{a b} \right]
\cda_{\bk a \si} c_{\bk b \si} + \mathcal{N} \hbar \omega_0 (\bda b + 1/2)
\nonumber \\
&+ \lambda \sum_{\bk, a, b, \si} C(\bk)_{a b}\cda_{\bk a \si} c_{\bk b \si} (\bda + b).
\end{align}
$\ep(\bk)_{a b}$ describe the dispersion in an orbital basis, and $\mu$ is the chemical potential.
$\cda_{\bk a \si}$ and $c_{\bk a \si}$ are electron creation
and annihilation operators, respectively, with lattice wavevector $\bk$, orbital index $a$, and spin $\si$.
The operators ($\bda$, $b$) describe
creation and annihilation operators for the $A_{1g}$ phonon with frequency $\om_0$, and $\mathcal{N}$ is the total number
of sites.
Electron-phonon interaction is described by $ \lambda C(\bk)_{a b}$, where $\lambda < 1$ is a
dimensionless small parameter and $C(k)_{ab}$ is order Fermi energy. Thus, electron-phonon interaction
can be treated perturbatively in orders of $\lambda$. For clarity, we ignore the phonon modes that
are not coherently generated. We also ignore electron-electron and phonon-phonon
interaction. Later, we comment on their effects.

After the pump the initial dynamics of the system is dominated by light-matter and by electron-electron interactions.
However, as time and angle resolved photoemission (tr-ARPES) experiments have
shown~\cite{papalazarou2012,yang2014}, due to electron-electron scattering the
electronic subsystem equilibrates after a time $\tau_r$ of order few tens of femtoseconds. At longer times an instantaneous
electronic temperature $T(t)$ can be defined.
In this work we focus on the regime $t \gg \tau_r$. Accordingly, we assume
$\tau_r \rightarrow 0$, such that the effect of the laser pump can be modeled as inducing a \emph{temperature quench} of the electrons.
We assume that the electronic temperature relaxation is characterized by a timescale $\tau_e$, and is described phenomenologically by
\beq
\label{eq:tau-T}
T(t) = T_L + (T_H - T_L) \operatorname{e}^{-t/\tau_e},
\eeq
where $T_L \equiv T(t=0^-) = T(t \rightarrow \infty)$, and $T_H \equiv T(t=0^+)$ [see Fig.~\ref{fig1}(b)].

The dimensionless mean atomic displacement $u \equiv \langle b + \bda \rangle$ follows the
equation of motion $\left( \partial_t^2  + \om_0^2 \right) u = F(t)$, where the out-of-equilibrium force is
\[
F(t) = - \frac{2 \om_0}{\mathcal{N}} \lambda \sum_{\bk, a, b, \si} C(\bk)_{a b}
\langle \cda_{\bk a \si} (t) c_{\bk b \si} (t) \rangle_{\ham, T(t)}.
\]
Here $\langle X \rangle_{\ham, T(t)} \equiv {\rm Tr} [\rho X]/{\rm Tr} [\rho]$ and $\rho \equiv |n \rangle  \langle n|
\operatorname{e}^{-E_n/T(t)}$, where $|n \rangle$ and $E_n$ are
the eigenfunctions and eigenvalues, respectively, of $\ham$ in Eq.~(\ref{eq:ham}).

Our goal is to capture, at least qualitatively, the feedback of the coherent phonon on the electron fluid,
for which it is sufficient to evaluate the force to second order in $\lambda$. At this order
$u(t)$ can be treated as a classical variable fluctuating in time, and $F(t)$ can be evaluated
using linear response theory. We get
\beq
\label{eq:force}
F(t)/(2 \om_0) = - \langle \hat{O} \rangle_{\ham_0, T(t)}
- \int_{- \infty}^{\infty} d \tp \Pi_{T(t)}(t - \tp) u(\tp),
\eeq
where $\Pi_{T(t)}(t - \tp) \equiv i \theta(t - \tp) \langle \left[ \hat{O} (\tp), \hat{O} (t) \right]
\rangle_{\ham_0, T(t)}$ is the response function associated with the weighted electron density operator
$\hat{O} \equiv (\lambda/\mathcal{N}) \sum_{\bk, a, b, \si} C(\bk)_{a b}\cda_{\bk a \si} c_{\bk b \si}$,
and $\ham_0 \equiv \ham(\lambda=0)$.
Since all the averages involving electronic operators from now on are
defined with respect to $\ham_0$, henceforth we do not mention it explicitly.
Note, as discussed in  Appendix~\ref{Appendix_structure}
$\Pi_{T(t)}(t - \tp)$ is a function not just of $(t - \tp)$, but also of $t$ via its dependence on temperature $T(t)$.
Moreover, the Fourier transform of the response function $\Pi_{T(t)}(\Om)$ coincides with the \emph{equilibrium}
retarded phonon self-energy $\Sigma_{\rm ph}(\Om)$ evaluated to second order in $\lambda$ and at temperature $T$ (see Eq.~(\ref{Ftrsf})).
At this stage it is also evident that, if needed, effects of electron-electron interaction can be systematically introduced
in the evaluation of $F(t)$.

The fact that the coherent phonon is a well-defined excitation implies that the retardation in $\Pi_{T(t)}(t - \tp)$
is weak, and it is sufficient to expand in frequency $\Pi_{T(t)}(\Om) \approx \pi(T) + i \Om \gamma(T)/\om_0$.
Here $\pi(T) \equiv \Pi_R(\Om=0, T)$ and $\gamma(T)/\om_0 \equiv \partial_{\Om}\Pi_I(\Om, T)_{\Om=0}$, where
$\Pi_{R/I}(\Om, T)$ are the real and imaginary parts of $\Pi_{T(t)}(\Om)$, respectively.
Note, in general, both $\pi(T)$ and $\gamma(T)$ are time dependent through their $T(t)$ dependencies. In the following
we simplify the discussion by assuming the decay rate $\gamma$ is constant, even though the
current formulation can handle time-dependent decay rates.
This gives
\beq
\label{eq:diff1}
\left( \partial_t^2  + 2 \gamma \partial_t + \om_0^2 \right) u = f(t),
\eeq
and
\[
f(t) \equiv - 2 \om_0 \left[ \langle \hat{O}  \rangle_{T}
-\langle \hat{O}  \rangle_{T_L} + \{ \pi(T) - \pi(T_L) \} u(t) \right]
\]
is the instantaneous out of equilibrium force.
In the above the second and the fourth terms are added by hand for the following reasons. The second term
involving $\langle \hat{O}  \rangle_{T_L}$ is a constant, and adding it is equivalent to setting
the zero of
the displacement $u$ to be the atomic position at $T_L$.
The fourth term involving $\pi(T_L)  u(t)$ renormalizes
the frequency $\om_0$ and adding it is equivalent to identifying
$\om_0$ with the equilibrium phonon frequency at $T_L$. Once these two terms are added,
we now get the behavior that is physically expected, namely $f(t=0^-) = f(t \rightarrow \infty) = 0$, see Eq.~(\ref{eq:tau-T}).

The functions $\langle \hat{O}  \rangle_{T}$ and $\pi(T)$ are well-defined thermodynamic quantities which, in the
absence of a phase transition, are analytic in $T$. Thus, they can be expanded around $T_L$ and,
using Eq.~(\ref{eq:tau-T}), they can
be expressed as series in powers of $\operatorname{e}^{- t/\tau_e}$.
In practice, these series can be truncated after the first few terms:
\begin{align}
\begin{aligned}
\langle \hat{O}  \rangle_{T} -\langle \hat{O}  \rangle_{T_L} = \sum_{n}a_{n}e^{-nt/\tau_e}
&\approx - (X_1/2) \operatorname{e}^{- t/\tau_1},
\\
\pi(T) - \pi(T_L) = \sum_{n}{}b_{n}e^{-nt/\tau_e} & \approx - (X_2/2) \operatorname{e}^{- t/\tau_2},
\end{aligned}
\label{eq:approx-equality}
\end{align}
where $T_L$ is the base temperature of pump-probe experiments,  $a_n = \left.\frac{d^{n}\langle \hat{O} \rangle_{T}}{dT^{n}}\right|_{T=T_L}(T_H-T_L)^{n}$,
 $b_n=\left.\frac{d^{n}\pi}{dT^{n}}\right|_{T=T_L}(T_H-T_L)^n$,
 $X_1 = -2 \left(\langle \hat{O}  \rangle_{T_H} -\langle \hat{O}  \rangle_{T_L} \right) \sim \mathcal{O}(\lambda)$,
$X_2 = -2 \left(  \pi(T_H) - \pi(T_L) \right) \sim \mathcal{O}(\lambda^2)$.
In other words, we assume that each of the series
$\sum_{n}a_{n}e^{-nt/\tau_e}$
and
$\sum_{n}{}b_{n}e^{-nt/\tau_e}$
can be modeled as a single decaying exponential
with effective decay rates $\ta_{1,2} \sim \tau_e$, respectively. The temperature dependencies of
$\langle \mathcal{O} \rangle_T$ and $\pi(T)$ can be obtained from the microscopic theory.
Then, the parameters  $[X_1, X_2, \tau_1, \tau_2]$ can be calculated using Eq. (5), provided
we know $[T_H, \tau_e]$. Hence \textit{the theory has only two phenomenological parameters},
namely $[T_H,\tau_e]$. We get
\beq
\label{eq:force1}
f(t) = \om_0 \left( X_1  \operatorname{e}^{- t/\tau_1} + u X_2 \operatorname{e}^{- t/\tau_2} \right),
\eeq
where the second term is the \emph{lattice feedback} which
can be interpreted as the effect of the change in the electron dispersion due to the coherent phonon excitation.
Eqs.~(\ref{eq:diff1}) and (\ref{eq:force1}), together with the initial conditions $u(0) =0$ and $\partial_t u(0) =0$, describe
the coherent phonon dynamics.

\section{Results}\label{Sec_results}
(i) Evaluating the force $f(t)$ to linear order in $\lambda$ is equivalent to ignoring the lattice feedback by
setting $X_2 =0$ in Eq.~(\ref{eq:force1}). In this limit, we recover
the phenomenological result of Zeiger \emph{et al.}~\cite{zeiger1992}, namely
$u(t) = (X_1/\om_0) [ \operatorname{e}^{- t/\tau_1} - \operatorname{e}^{- \gamma t} \cos (\om_0 t - \phi_0)/ \cos \phi_0 ]$,
with the phase $\phi_0 \sim {\rm max}[\gamma/\om_0, 1/(\om_0 \tau_1)]$.
However, the detection of a coherent phonon necessarily implies that
in a typical experimental situation
\beq
\label{eq:inequality}
\om_0 \gg \gamma, 1/\tau_{1/2},
\eeq
and so $\phi_0 \ll 1$, which means that the phase obtained within the phenomenological framework is negligible.
As we show below, keeping the lattice feedback term also leads to a finite phase of a different physical
origin, and this latter is quantitatively
more significant than $\phi_0$.

(ii) Finite $X_2$ leads to a richer dynamics and a modified solution.
In the limit $[\gamma/\om_0, 1/(\om_0 \tau_{1/2})] \rightarrow 0$, which is
experimentally relevant, we get (see Eq.~(\ref{u_sol}))
\beq
\label{eq:full-soln}
u(t) = \frac{X_1 \operatorname{e}^{- t/\tau_1}}{\om_0 - X_2 \operatorname{e}^{- t/\tau_2}}
- \frac{X_1 \operatorname{e}^{- \gamma t} }{\om_0 - X_2} \cos [\om_0 t + \Phi(t)],
\eeq
where
\beq
\label{eq:phase}
\Phi(t) \equiv - \frac{X_2 \tau_2}{2} \left( 1 - \operatorname{e}^{- t/\tau_2} \right).
\eeq
Equations~(\ref{eq:full-soln}) and (\ref{eq:phase}) summarize the main results of this work.
%====================
\begin{figure}[!!t]
\begin{center}
\includegraphics[width=1.0\linewidth,trim=0 0 0 0]{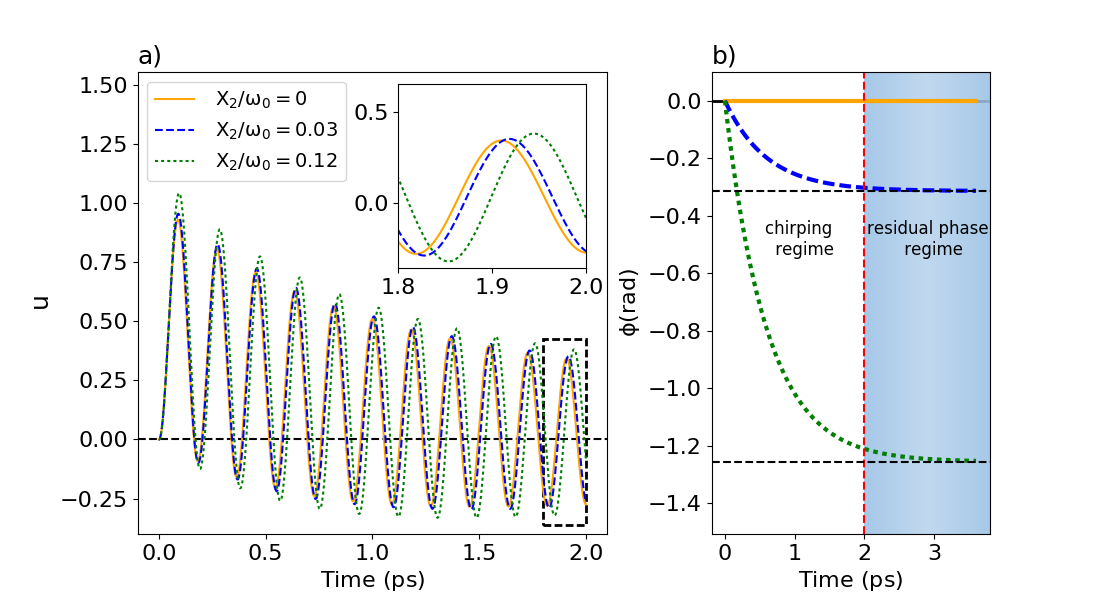}
\caption{
(color online)
Calculations for representative parameter values.
Frequency $\om_0/(2\pi) =5.5$ THz, $X_1/\om_0 = 0.5$, $\tau_1 = 0.7$ ps,
$\tau_2 = 0.6$ ps, and $\gamma^{-1} =5$ ps, and for
different strengths of the lattice feedback term $X_2$.
$X_2=0$ corresponds to the phenomenological theory~\cite{zeiger1992}.
(a) Coherent phonon displacement $u(t)$, see Eqs.~(\ref{eq:full-soln}) and (\ref{eq:phase}) and the associated text.
The inset, a blow-up of the
dashed rectangle, shows signature of the finite phase $\phi$ for different values of $X_2$.
(b) The effects of the feedback at different time scales.
At short times ($t \lesssim \tau_2$) a finite
$X_2$ leads to \emph{chirping}. At long times ($t \gg \tau_2$) it leads to a finite phase $\phi$,
see also inset in (a).
$\tau_2$ is defined in Eq.~(\ref{eq:approx-equality}).
}
\label{fig2}
\end{center}
\end{figure}
%====================

At face value, the above is a five parameter description of the coherent phonon. However, if the microscopic
prescription is followed, $(X_1, X_2, \tau_1, \tau_2)$ can be obtained from the phenomenological
parameters $T_H$ and $\tau_e$ defined in Eq.~(\ref{eq:tau-T}) by using the approximate relations
of Eq.~(\ref{eq:approx-equality}).
Furthermore, if the theory to $\mathcal{O}(\lambda^2)$ is quantitatively
sufficient, then $\gamma^{-1}$ is the equilibrium phonon lifetime measured by, say, Raman response.

(iii) For $t \lesssim \tau_2$ the feedback $\Phi(t)$ describes temporal variation of the oscillation
frequency, i.e., \emph{chirping}, with a frequency variation $\Delta \om_0 \sim -X_2/2$, see Fig.~\ref{fig2}.
On the other hand, for $t \gg \tau_2$ we get a finite residual phase
$\phi \equiv \Phi(t \rightarrow \infty) = -X_2 \tau_2/2$, see Fig.~\ref{fig2}.
Note, even if $\left|\Delta \om_0 \right|/ \om_0 \ll 1$ and the chirping is not experimentally observable at low fluence, the
phase $\phi = (\Delta \om_0/ \om_0)(\om_0 \tau_2)$ can be substantial since it involves the large parameter $\om_0 \tau_2$,
c.f., Eq.~(\ref{eq:inequality}). Note, the time dependent phase $\Phi(t)$ is qualitatively different from a constant
phase that is usually discussed in the literature.

The chirping discussed here is related to the temperature, and hence, to the time dependence of the phonon frequency
due to electron-phonon interaction. This is to be contrasted with other mechanisms of
chirping discussed in the literature such as that due to phonon anharmonicity~\cite{hase2002} and carrier
diffusion~\cite{tangney1999,tangney2002,fritz2007}.

(iv) Equilibrium Raman spectroscopy of BaFe$_2$As$_2$ shows that the
$A_{1g}$ phonon frequency softens with increasing temperature~\cite{rahlenbeck2009}.
Simultaneously, the phonon lifetime~\cite{rahlenbeck2009} has an atypical temperature dependence
across the magnetic transition of BaFe$_{2}$As$_{2}$ which is very reminiscent of the $T$-dependence of
resistivity~\cite{rullier-albenque}, implying that the phonon temperature dependencies
are likely due to interaction with the electrons.
Thus, from these equilibrium trends, we conclude that
$X_2 > 0$, and we \emph{predict} that the coherent $A_{1g}$ phonon of BaFe$_2$As$_2$ will show red-shifted chirp
at sufficiently high fluence.

(v) Since in our theory the frequency shift $\Delta \om_0$ and the residual phase $\phi$ both depend on $X_2$,
an important conclusion is that red-shifted (blue shifted) chirp is accompanied by negative (positive)
residual phase. Note, the above expectation is indeed correct for the $A_{1g}$ coherent phonon of
BaFe$_2$As$_2$, which softens with increasing temperature, and for which a negative phase $\phi = - 0.1\pi$ has been
reported~\cite{yang2014,rettig2015}, see also the discussion in Sec.~\ref{Ba_model}.

\section{Quantitative description of the $A_{1g}$ coherent phonon in BaFe$_{2}$As$_{2}$  }
\label{Ba_model}

\begin{figure}[!!t]cc
\begin{center}
\includegraphics[width=1.0\linewidth,trim=0 0 0 0]{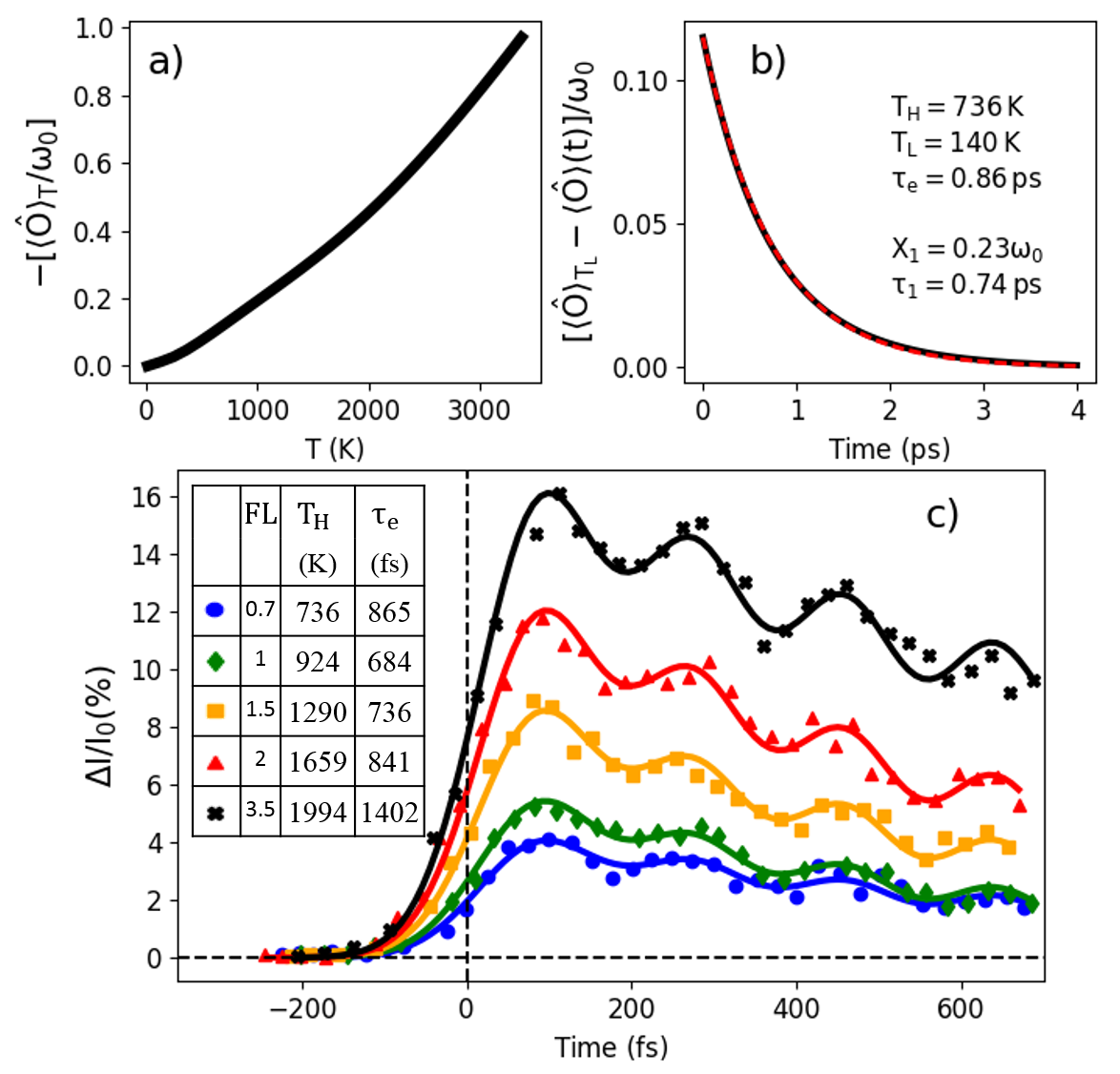}
\caption{
(color online) Quantitative description of the $A_{1g}$ coherent phonon (frequency $\om_0/(2\pi) =5.5$ THz) of BaFe$_2$As$_2$,
and comparison with experiment~\cite{rettig2015}.
(a) Calculated equilibrium expectation value of the weighted electron density $\langle \hat{O}  \rangle_{T}$
for $\lambda =$ 0.25.
(b) Solid (black) line: The $T$-dependence in (a) is transformed into a time dependence using Eq.~(\ref{eq:tau-T})
for representative
values of the phenomenological parameters $(T_H,\tau_e)$. Base temperature $T_L = 140$ K. Dashed (red) line:
Fit using Eq.~(\ref{eq:approx-equality}), and estimate of $(X_1, \tau_1)$.
(c) Solid lines: temporal variation of x-ray form factor calculated using Eq.~(\ref{eq:full-soln})
at different fluences (FL in mJ/cm$^2)$.
The table gives estimates of $(T_H,\tau_e)$ used in the calculation. The fit uses $\gamma^{-1} = 5$ ps, which
is the equilibrium lifetime~\cite{rahlenbeck2009}.
Symbols represent data points extracted from Ref.~\cite{rettig2015}.
}
\label{fig3}
\end{center}
\end{figure}

In this section, we apply the theory quantitatively to the coherent $A_{1g}$ phonon of the strongly correlated
metal BaFe$_2$As$_2$, and we compare the theory results with a recent time-resolved x-ray study~\cite{rettig2015},
see Fig~\ref{fig3}.
BaFe$_2$As$_2$ is the parent compound of a class of high temperature superconductors that also have rather
interesting magnetic and nematic properties \cite{Johnston}. The $A_{1g}$ coherent phonon in this system, associated with the
motion of the As atoms, has also been widely studied using a variety of pump-probe techniques \cite{yang2014,rettig2015,kim2012,mansart2010}, including
time-resolved x-ray spetcroscopy \cite{rettig2015}which provides the most direct information about the As motion.
The electronic properties of the BaFe$_2$As$_2$  are known \cite{Egami} to be very sensitive to the As height, which makes
the study of the coherent phonon motion all the more interesting.

Our overall goal in this section is to check to what extent a microscopic tight-binding model, that has been successfully
used to understand equilibrium properties, can be used to describe the transient temperature dependencies involved
in a pump-probe setting. Such an exercise is a step in the direction of extracting information about equilibrium
properties from a pump-probe setup.

As a first, step we define the various parameters that we use to describe
BaFe$_2$As$_2$ with the microscopic Hamiltonian of Eq.~(\ref{eq:ham}).
We take the electronic kinetic part $\epsilon_{ab}(\textbf{k})$ from  Ref.~\cite{graser2010}, which itself
is obtained as a tight-binding fit of the LDA band structure onto a basis of five $d$ Fe orbitals~\cite{Unfold}.
Note, this particular set of tight-binding parameters has been
used widely in the literature. Relatively less detailed information is currently available concerning the orbitally
resolved electron-phonon matrix elements $C(\textbf{k})_{ab}$ of Eq.~(\ref{eq:ham}). However, it is well-accepted that
an increase of the dimensionless arsenic height $u=\langle b^{\dagger}+b \rangle$ is accompanied by a reduction of
the hopping-integrals and the bandwidths~\cite{kuroki2009} since the hopping of the electrons between Fe atoms can also
be mediated by the As atoms.
Taking into account this physical expectation, we found that a simple way to model the electron-phonon matrix elements is to assume
\begin{equation}
    C(\bk)_{a b}= -[t_{nn}]_{a b}(\textbf{k}),
\end{equation}
where  $[t_{nn}]_{ab}$ is the diagonal nearest-neighbour entries of the tight-binding parameters $\epsilon_{ab}(\textbf{k})$.
Thus, in our scheme the entire electron-phonon coupling is ultimately described by a single additional dimensionless parameter
$\lambda$ which can later be absorbed in an overall scaling factor between the calculated $u(t)$ and the experimental x-ray intensity
(see also the discussion following Eq.~(\ref{SFit}) below).

As a second step, we describe the calculation of the out-of-equilibrium force $F(t)$ (see also the discussion in the paragraph
following Eq.~(\ref{eq:tau-T})) to first order in $\lambda$.
This involves the calculation of the thermal average of the weighted electron
density operator. From Eq.~(\ref{eq:force}) we get
\begin{equation}\label{Sforcing}
\begin{split}
\langle \hat{O}\rangle_{\ham_0,T} & \equiv  \frac{\lambda}{\mathcal{N}}
\sum_{\bk, a, b, \si} C(\bk)_{a b}\langle \cda_{\bk a \si}  c_{\bk b \si}  \rangle_{\ham_0, T} \\
& =\frac{\lambda}{\mathcal{N}}
\sum_{\bk, \nu, \si} \tilde{C}(\bk)_{\nu \nu} n_F\big[\xi_{\nu}(\textbf{k})-\mu(T),T\big],
\end{split}
\end{equation}
where the last equality is written in the band basis.
Here $n_F$ is the Fermi function, $\xi_{\nu}(\textbf{k})$ is the energy of an electron in the
band $\nu$ with momentum $\textbf{k}$, $\tilde{C}(\bk)_{\nu \nu}$ is the electron-phonon matrix
elements in the band basis, and $\mu(T)$ is the chemical potential at the transient temperature $T(t)$ at time $t$.
We assume that there is no electronic diffusion~\cite{torchinsky2011}, and that the particle number
is conserved during the pump-probe cycle, which is consistent with the conclusions of a
recent time-resolved photoemission study \cite{yang2014}.
We divide the Brillouin zone into a ($10\times10\times10$) grid,
and diagonalize $\ham_0$ at each point of the grid to obtain the electronic dispersion $\xi_{\nu}(\textbf{k})$.
The chemical potential is then calculated by solving the particle number conservation equation numerically.
In Fig.~\ref{fig3} (a) we show the result of our calculation of
$\langle \hat{O}\rangle_{\ham_0,T}$
for temperatures ranging from 0 to 3500 $\rm{(K)}$. This $T$-dependence can be
transformed into a time dependence using Eq.~(\ref{eq:tau-T})  provided we have an estimate
of the  phenomenological parameters $(T_H,\tau_e)$ at each pump fluence.
Henceforth, the base temperature is taken as $T_L = 140$ K.
The solid (black) line of Fig.~\ref{fig3} (b) gives such a transformation
$\langle \hat{O}\rangle_{\ham_0,T} \rightarrow \langle \hat{O}\rangle (t)$ for a representative
value of $(T_H,\tau_e)$.
The resulting time dependence can be modeled by a
single decaying exponential using Eq.~(\ref{eq:approx-equality}). This leads to an estimate of $(X_1, \tau_1)$
for each pump fluence, see dashed (red) line of Fig.~\ref{fig3} (b).
%====================
%====================

Note, the above step should not be construed as a mere replacement of two phenomenological parameters
$(X_1, \tau_1)$ by two other phenomenological parameters $(T_H,\tau_e)$. This is because in our scheme
the estimation of $(X_1, \tau_1)$ at each fluence is obtained via
the evaluation of $\langle \hat{O}\rangle_{\ham_0,T}$ from
the microscopic Hamiltonian Eq.~(\ref{eq:ham}) whose parameters are themselves fluence independent.
Thus, the modeling is highly constrained, and it is not obvious that the $(X_1, \tau_1)$ needed for a
given fluence can be obtained in our scheme for reasonable values of $(T_H,\tau_e)$ once the Hamiltonian is
fixed.
One way to
appreciate the nontrivial step involved in our quantitative modeling is to note
that our scheme can provide meaningful $(T_H,\tau_e)$ only if $\langle \hat{O}\rangle_{\ham_0,T}$
is a \emph{monotonically increasing} function of temperature. On the other hand, such a property is \emph{a priori}
not guaranteed. Likewise, if the slope of the function
$\langle \hat{O}\rangle_{\ham_0,T}$ is too large/small it would lead to values of $T_H$ that are too small/large
compared to the estimates currently available from time-resolved photoemission studies \cite{yang2014}.

In the third step we discuss the relevance of the $\lambda^2$ contribution to the force $F(t)$ that is implied in the
experiment of Ref.~\cite{rettig2015,yang2014}. This contribution can be estimated from the following argument.
To $\lambda^2$ accuracy,  $\pi(T)$ can also be identified as the equilibrium
phonon self-energy whose $T$-dependence can be inferred from equilibrium Raman measurement of $\om_0(T)$~\cite{rahlenbeck2009}.
For $T_L = 140$ K and $T_H \sim 500$ K,
an extrapolation of $\om_0(T)$ reported in Ref.~\cite{rahlenbeck2009} gives $\Delta \omega_0 = 0.4$ THz, and therefore
$\frac{X_2}{\omega_0} \approx 0.01$, see Eq.~(\ref{eq:phase}).
This small fraction implies that the $\lambda^2$ contribution to the force $F(t)$ is unimportant for the fluences used
in Ref.~\cite{rettig2015}. Nevertheless, for the fits we kept the phase $\Phi(t)$ generated by the feedback effect, and we used the
expression
\beq
\label{eq:fit-u}
u(t) = (X_1/\om_0) (\operatorname{e}^{- t/\tau_1}
- \operatorname{e}^{- \gamma t} \cos [\om_0 t + \Phi(t)]),
\eeq
by setting $\frac{X_2}{\omega_0} \rightarrow$ 0 in Eq.~(\ref{eq:full-soln}).
To model $\Phi(t)$ we assume that it is fluence independent and that the experimentally
reported phase $\phi = - 0.1 \pi$~\cite{rettig2015,yang2014} can be identified with $\Phi(t \rightarrow \infty) = -X_2 \tau_2/2$
(see Eq.~(\ref{eq:phase})), from which we get $\tau_2 \approx 800\,\rm{fs}$.
Note also, for time $t\lesssim \tau_e$ the quality of the fit is marginally
affected by including the feedback $\Phi(t)$ term.

Thus, following the above three steps we are able to compute $u(t)$ for a given
fluence provided we have an estimate of $(T_H,\tau_e)$.

Finally, we compare the calculated arsenic displacement $u(t)$ with that measured in time resolved
x-ray scattering~\cite{rettig2015} for
a fluence range of $0.7$ to $3.5\,\rm{(mJ/cm^2)}$.
The intensity is convolved with a Gaussian pulse to account for the limited time resolution~\cite{rettig2015}.
In the kinematic approximation~\cite{rettig2015}
the variation of the intensity is proportional to the arsenic displacement and is given by
\beq\label{SFit}
\frac{\Delta I}{I_{0}}(t)=  \frac{B}{ \tau_r\sqrt{\pi} }\int^{\infty}_0 e^{-(\frac{t-\tau}{\tau_r})^2}
u(\tau) d\tau,
\eeq
where
$I_0$ is the equilibrium intensity, $\Delta I$ is the variation of intensity
out of equilibrium, $\tau_r\approx 96 \,\rm{fs}$ is the experimental resolution of the probe-pulse,
and $u(t)$ is computed using Eq.~(\ref{eq:fit-u}) following the three steps mentioned above.
$B$ is a dimensionless proportionality constant, independent of fluence, that sets the overall scale of the
theoretically evaluated $\Delta I/I_0$ with respect to the experimentally measured ones. Physically, $B$ is
related to the change of the relevant x-ray form factor with the As atomic position.
Within our scheme the constant $B$ and the dimensionless electron-phonon coupling $\lambda$ cannot be
estimated separately. We find that best fits are obtained for $\lambda B =$ 4.9.
In Fig.~\ref{fig3} (c) we compare the calculated
$\Delta I/I_0$ (lines) with the data of Ref.~\cite{rettig2015} (solid symbols).

From Fig.~\ref{fig3}(c)
we conclude that the two-parameter fit is quite reasonable, given the simplicity of the starting model.
Furthermore, our
estimation of $(T_H, \tau_e)$, given in the inset of Fig.~\ref{fig3} (c), compares well with the
experimental estimations given in Ref.~\cite{yang2014}.
The above attempt at a quantitative description is an important step towards making
connection between equilibrium microscopic description
of electrons with out-of-equilibrium pump-probe data. Note, the above calculation does not include
temperature dependencies of the single electron properties arising due to
electron-electron interaction.
While such interaction effects can be
incorporated in the current formalism, it is beyond the scope of the current work.

\section{Conclusions} \label{conclusion}
We developed a microscopic theory of displacive coherent phonons driven by laser heating of carriers.
Our theory captures physics beyond the standard phenomenological description, namely
the modification of the electronic energy levels due to the phonon excitation, and how this
change feeds back on the phonon dynamics. This effect of electron-phonon interaction leads to chirping at short time scales,
and at long times it appears as a finite phase in the oscillatory signal.
We successfully applied the theory to the $A_{1g}$ coherent phonon of BaFe$_2$As$_2$,
thereby demonstrating that pump-probe data can be related to microscopic quantities and eventually to
equilibrium physics.
We explained the origin of the phase in the oscillatory signal reported in recent experiments~\cite{yang2014,rettig2015}
on this system, and we predict
that it will exhibit red-shifted chirping at larger fuence.

\acknowledgments
We are thankful to M. Bauer, V. Brouet, I. Eremin,
Y. Gallais, L. Perfetti, M. Schiro, K. Sengupta, A. Subedi for insightful discussions.
We acknowledge financial support from ANR grant ``IRONIC'' (ANR-15-CE30-0025).

\appendix

\section{Structure of $\Pi_{T(t)}(t - \tp)$}\label{Appendix_structure}

The response function used in the main text  is defined by
\beq
\Pi_{T(t)}(t - \tp) \equiv i \theta(t - \tp) \langle \left[ \hat{O} (\tp), \hat{O} (t) \right]
\rangle_{\ham_0, T(t)},
\eeq
where $\hat{O} \equiv (\lambda/\mathcal{N}) \sum_{\bk, a, b, \si} C(\bk)_{a b}\cda_{\bk a \si} c_{\bk b \si}$
is the weighted electron density operator,
$\ham_0 \equiv \ham(\lambda=0)$, and the Hamiltonian $\ham$ is given by Eq.~(1) in the main text. In equilibrium
$\Pi_{T(t)}(t - \tp)$ is a function of $(t-\tp)$ only, but this is no longer the case out-of-equilibrium. Here,
we discuss the $t$ and $\tp$ dependencies of $\Pi_{T(t)}(t - \tp)$. We write the response function in the Lehmann
representation where the time structure can be made explicit
\begin{equation}\label{Lehmann_rep}
\begin{split}
\Pi_{T(t)}(t - \tp) & = i\theta(t-t')\sum_{n,m} \lvert\bra{n}\hat{O}\ket{m}\rvert^2 e^{i(t-t')(E_{n}-E_{m})} \\
& \times \bigg( e^{-\beta(t)E_{m}} - e^{-\beta(t)E_{n}}\bigg),
\end{split}
\end{equation}
where  $E_{n}$ and $\ket{n}$ are respectively a complete set of eigenenergies  and eigenstates of the Hamiltonian
$\ham_0$. We see from (\ref{Lehmann_rep}) that the response function is a function of the time difference ($t-\tp$),
and that the explicit time $t$ dependence  enters only through the electronic temperature $T(t)$. We can then define
the Fourier transform of the response function with respect to the time difference ($t-\tp$) evaluated at the electronic
temperature $T(t)$
\begin{equation}\label{Ftrsf}
\begin{split}
\Pi_{T(t)}(\Omega) &=  \sum_{n,m} \lvert\bra{n}\hat{O}\ket{m}\rvert^2 \frac{1}{E_{n}-E_{m}-\Omega+i\eta }\\ &\times
\bigg( e^{-\beta(t)E_{n}} - e^{-\beta(t)E_{m}}\bigg),
\end{split}
\end{equation}
where $\eta$ is an arbitrarily small positive constant that ensures the convergence of the Fourier transform.
By inspection, we see that the response function in frequency domain (\ref{Ftrsf}) is the equilibrium retarded phonon
self-energy evaluated to second order in the electron-phonon interaction ($\lambda^2$) at temperature $T(t)=1/\beta(t)$.

\section{Solution of the differential equation for $u(t)$}\label{Appendix_solution_u}

As discussed in the main text, if the instantaneous out-of-equilibrium force $f(t)$ is evaluated to second order in
electron-phonon interaction the theory captures the modification of the electronic dispersion due to the coherent
phonon excitation, and how that feeds back upon the dynamics of the phonon itself. Taking this feedback into account
the differential equation governing the atomic displacement $u(t)$ is given by [see Eqs.~(4) and (6) in main text]
\beq \label{Eq_Coh}
\left( \partial_t^2  + 2 \gamma \partial_t + \om_0^2 \right) u = f(t) =
\om_0 \left( X_1 \operatorname{e}^{- t/\tau_1} + u X_2 \operatorname{e}^{- t/\tau_2} \right).
\eeq
The parameters $(\om_0, X_1, X_2, \gamma, \tau_{1/2})$ are defined in the main text. Here we discuss the solution of
the above differential equation subject to the initial conditions $u(0) =0$ and $\partial_t u(0) =0$, and in the
experimentally relevant limit of $[\gamma/\om_0, 1/(\om_0 \tau_{1/2})] \rightarrow 0$. The equation of motion (\ref{Eq_Coh})
is linear, the solution is then the sum of the homogeneous and particular solution $u(t)=y_h(t)+y_p(t)$. We first discuss the
homogeneous solution, then following the same method we give the particular solution. We start from the following ansatz
for the homogeneous solution
\beq\label{yh_Form}
\begin{split}
y_h(t) &= \sum_{n=0}^{\infty} a_n e^{k_n t} + cc, \\
\end{split}
\eeq
with $k_n= i \omega_1  - \gamma  -  n/\tau_2$, and $\omega_1=\sqrt{(\omega_0)^2-\gamma^2}$. We insert (\ref{yh_Form}) into
the homogeneous equation, and obtain an equation for the coefficients  $a_n$
\beq\label{Coef_eq}
\begin{split}
 a_n( k_n^2 + 2\gamma_0 k_n + (\omega_0)^2) &= \omega_0 X_2 a_{n-1}, \\
 a_0( k_0^2 + 2\gamma_0 k_0 + (\omega_0)^2) &= 0.
\end{split}
\eeq
Since $k_0$ satisfies the equation $( k_0^2 + 2\gamma_0 k_0 + (\omega_0)^2)=0$, $a_0$ is then an arbitrary complex  constant.
We solve the coupled equation (\ref{Coef_eq}), and obtain for $a_n$
\beq
\begin{split}
a_n&=a_0( \omega_0 X_2 )^n \prod_{m=1}^{n} \frac{1}{ k_m^2+2\gamma k_m + \omega_0^2}\\
&=\frac{a_0}{n!}\left(\frac{i X_2 \tau_2 }{2\omega_1}\right)^n \prod_{m=1}^{n} \frac{1}{  1 + (2 i)(m/\tau_2\omega_0)} \\
&\approx \frac{a_0}{n!}\left(\frac{i X_2 \tau_2 }{2\omega_0}\right)^n,
\end{split}
\eeq
where in the last step we took the limit $[\gamma/\om_0, 1/(\om_0 \tau_{2})] \rightarrow 0$, the homogeneous solution then reads
\beq
\begin{split}
y_h(t) &= a_0 e^{(i\omega_0 - \gamma) t}\sum_n^{\infty} \frac{1}{n!}\left(\frac{i X_2 \tau_1 }
{2\omega_0} e^{-t/\tau_2}\right)^n+ c.c. \\
&= a_0 e^{-\gamma t}\bigg(e^{i\omega_0+i\frac{ X_2 \tau_2 }{2\omega_0} e^{-t/\tau_2}}\bigg)+ c.c. ,
\end{split}
\eeq
We replace $a_0=\frac{1}{2}Ae^{i \psi}$ and finally obtain for the homogeneous solution
\beq\label{y_h_solution}
y_h(t) =  A e^{-\gamma t} \cos{\bigg(\omega_0 t + \frac{X_2\tau_2}{2\omega_0}e^{-t/\tau_2} + \psi\bigg)},
\eeq
where $(A,\psi)$ are arbitrary constants to be determined from the initial conditions. We follow the same method to find
the particular solution, we start from the ansatz
\beq\label{yp_Form}
y_p(t) = \sum_{n=0}^{\infty} b_n e^{\alpha_n t}, \\
\eeq
with $\alpha_n= - 1/\tau_1  -  n/\tau_2$. We insert (\ref{yp_Form}) into the equation of motion (\ref{Eq_Coh}),
and get an equation for the coefficients $b_n$
\beq
\begin{split}
b_n( \alpha_n^2 + 2\gamma_0 \alpha_n + (\omega_0)^2) &= \omega_0 X_2 a_{n-1},\\  b_0( \alpha_0^2
+ 2\gamma_0 \alpha_0 + (\omega_0)^2) & = \omega_0 X_1.
\end{split}
\eeq
We solve the coupled equations and obtain
\beq
\begin{split}
b_n  &=\frac{\omega_0 X_1 (\omega_0 X_2)^n }{(1/\tau_1)^2-2\gamma(1/\tau_1)+(\omega_0)^2} \prod_{m=1}^{n}
\frac{1}{  \alpha_m^2 +2\gamma \alpha_m+\omega_0^2} \\
&\approx \frac{X_1}{\omega_0} \left(\frac{ X_2  }{\omega_0}\right)^n,
\end{split}
\eeq
where in the last step we took the limit $[\gamma/\om_0, 1/(\om_0 \tau_{1/2})] \rightarrow 0$, the particular solution then reads
\beq
\begin{split}
y_p(t) &= \frac{X_1}{\omega_0} e^{-t/\tau_1}\sum_n^{\infty} \left(\frac{ X_2 e^{-t/\tau_2}  }{\omega_0}\right)^n \\
& = \frac{X_1}{\omega_0-X_2e^{-t/\tau_2}} e^{-t/\tau_1}.
\end{split}
\eeq
We use the initial conditions $u(0)=0$ and $\partial_t u (0)=0$ to calculate the arbitrary constants $(A,\psi)$.
The solution in the limit $[\gamma/\om_0, 1/(\om_0 \tau_{1/2})] \rightarrow 0$ reads
\beq\label{u_sol}
\begin{split}
u(t)& =  \frac{X_1}{\omega_0-X_2e^{-t/\tau_2}} e^{-t/\tau_1}\\ & -\frac{X_1}{\omega_0-X_2} e^{-\gamma t}
\cos{\bigg(\omega_0 t + \frac{X_2\tau_2}{2\omega_0}\big(e^{-t/\tau_2}-1\big) \bigg)},
\end{split}
\eeq
where we finally recognize Eq~(8) of the main text.

\end{document}